\input harvmac

\input amssym
\input epsf

\def\unit{\relax{\rm 1\kern-.26em I}}
\def\nada{\relax{\rm 0\kern-.30em l}}
\def\tilde{\widetilde}

\def\alphadot{{\dot \alpha}}



\noblackbox
\def\IL{\relax{\rm I\kern-.18em L}}
\def\IH{\relax{\rm I\kern-.18em H}}
\def\IR{\relax{\rm I\kern-.18em R}}
\def\IC{\relax\hbox{$\inbar\kern-.3em{\rm C}$}}
\def\IZ{\relax\ifmmode\mathchoice
{\hbox{\cmss Z\kern-.4em Z}}{\hbox{\cmss Z\kern-.4em Z}} {\lower.9pt\hbox{\cmsss Z\kern-.4em Z}}
{\lower1.2pt\hbox{\cmsss Z\kern-.4em Z}}\else{\cmss Z\kern-.4em Z}\fi}
\def\CM {{\cal M}}

\def\partialslash{\not{\hbox{\kern-2pt $\partial$}}}

\def\CL {{\cal L}}

\def\CO {{\cal O}}

\def\CM {{\cal M}}

\def\CO {{\cal O}}

\font\manual=manfnt \def\dbend{\lower3.5pt\hbox{\manual\char127}}

\def\IZ{\relax\ifmmode\mathchoice
{\hbox{\cmss Z\kern-.4em Z}}{\hbox{\cmss Z\kern-.4em Z}} {\lower.9pt\hbox{\cmsss Z\kern-.4em Z}}
{\lower1.2pt\hbox{\cmsss Z\kern-.4em Z}}\else{\cmss Z\kern-.4em Z}\fi}
\def\half {{1\over 2}}

\def\bar{\overline}

\def\rt2{\sqrt{2}}
\def\irt2{{1\over\sqrt{2}}}

\def\slashchar#1{\setbox0=\hbox{$#1$}           
   \dimen0=\wd0                                 
   \setbox1=\hbox{/} \dimen1=\wd1               
   \ifdim\dimen0>\dimen1                        
      \rlap{\hbox to \dimen0{\hfil/\hfil}}      
      #1                                        
   \else                                        
      \rlap{\hbox to \dimen1{\hfil$#1$\hfil}}   
      /                                         
   \fi}

\def\foursqr#1#2{{\vcenter{\vbox{
    \hrule height.#2pt
    \hbox{\vrule width.#2pt height#1pt \kern#1pt
    \vrule width.#2pt}
    \hrule height.#2pt
    \hrule height.#2pt
    \hbox{\vrule width.#2pt height#1pt \kern#1pt
    \vrule width.#2pt}
    \hrule height.#2pt
        \hrule height.#2pt
    \hbox{\vrule width.#2pt height#1pt \kern#1pt
    \vrule width.#2pt}
    \hrule height.#2pt
        \hrule height.#2pt
    \hbox{\vrule width.#2pt height#1pt \kern#1pt
    \vrule width.#2pt}
    \hrule height.#2pt}}}}
\def\psqr#1#2{{\vcenter{\vbox{\hrule height.#2pt
    \hbox{\vrule width.#2pt height#1pt \kern#1pt
    \vrule width.#2pt}
    \hrule height.#2pt \hrule height.#2pt
    \hbox{\vrule width.#2pt height#1pt \kern#1pt
    \vrule width.#2pt}
    \hrule height.#2pt}}}}
\def\sqr#1#2{{\vcenter{\vbox{\hrule height.#2pt
    \hbox{\vrule width.#2pt height#1pt \kern#1pt
    \vrule width.#2pt}
    \hrule height.#2pt}}}}

\def\figin{\epsfcheck\figin}\def\figins{\epsfcheck\figins}
\def\epsfcheck{\ifx\epsfbox\UnDeFiNeD
\message{(NO epsf.tex, FIGURES WILL BE IGNORED)}
\gdef\figin##1{\vskip2in}\gdef\figins##1{\hskip.5in}
\else\message{(FIGURES WILL BE INCLUDED)}%
\gdef\figin##1{##1}\gdef\figins##1{##1}\fi}
\def\DefWarn#1{}
\def\figinsert{\goodbreak\midinsert}
\def\ifig#1#2#3{\DefWarn#1\xdef#1{fig.~\the\figno}
\writedef{#1\leftbracket fig.\noexpand~\the\figno}%
\figinsert\figin{\centerline{#3}}\medskip\centerline{\vbox{\baselineskip12pt \advance\hsize by
-1truein\noindent\footnotefont{\bf Fig.~\the\figno:\ } \it#2}}
\bigskip\endinsert\global\advance\figno by1}


\lref\CheungQF{
  C.~Cheung, J.~Mardon, Y.~Nomura and J.~Thaler,
  ``A Definitive Signal of Multiple Supersymmetry Breaking,''
  JHEP {\bf 1007}, 035 (2010)
  [arXiv:1004.4637 [hep-ph]].
}

\lref\WessCP{
  J.~Wess and J.~Bagger,
  ``Supersymmetry and supergravity,''
{\it  Princeton, USA: Univ. Pr. (1992) 259 p}
}

\lref\BaggerGM{
  J.~A.~Bagger and A.~F.~Falk,
  ``Decoupling and Destabilizing in Spontaneously Broken Supersymmetry,''
  Phys.\ Rev.\  D {\bf 76}, 105026 (2007)
  [arXiv:0708.3364 [hep-ph]].
}

\lref\RocekNB{
  M.~Rocek,
  ``Linearizing The Volkov-Akulov Model,''
  Phys.\ Rev.\ Lett.\  {\bf 41}, 451 (1978).
}

\lref\VolkovIX{
  D.~V.~Volkov and V.~P.~Akulov,
  ``Is the Neutrino a Goldstone Particle?,''
  Phys.\ Lett.\  B {\bf 46}, 109 (1973).
}

\lref\SamuelUH{
  S.~Samuel and J.~Wess,
  ``A Superfield Formulation Of The Nonlinear Realization Of Supersymmetry And
  Its Coupling To Supergravity,''
  Nucl.\ Phys.\  B {\bf 221}, 153 (1983).
}

\lref\nonlinear{
  Z.~Komargodski and N.~Seiberg,
  ``From Linear SUSY to Constrained Superfields,''
  arXiv:0907.2441 [hep-th].
}

\lref\KuzenkoEF{
  S.~M.~Kuzenko and S.~J.~Tyler,
  ``Relating the Komargodski-Seiberg and Akulov-Volkov actions: Exact nonlinear
  field redefinition,''
  arXiv:1009.3298 [hep-th].
}

\lref\ZheltukhinXR{
  A.~A.~Zheltukhin,
  ``On equivalence of the Komargodski-Seiberg action to the Volkov-Akulov
  action,''
  arXiv:1009.2166 [hep-th].
}

\lref\LiuSK{
  H.~Liu, H.~Luo, M.~Luo and L.~Wang,
  ``Leading Order Actions of Goldstino Fields,''
  arXiv:1005.0231 [hep-th].
}

\lref\CraigYF{
  N.~Craig, J.~March-Russell and M.~McCullough,
  ``The Goldstini Variations,''
  JHEP {\bf 1010}, 095 (2010)
  [arXiv:1007.1239 [hep-ph]].
}

\lref\KomargodskiRB{
  Z.~Komargodski and N.~Seiberg,
  ``Comments on Supercurrent Multiplets, Supersymmetric Field Theories and
  Supergravity,''
  JHEP {\bf 1007}, 017 (2010)
  [arXiv:1002.2228 [hep-th]].
}

\lref\Brignole{
A.~Brignole,
  ``One-loop Kaehler potential in non-renormalizable theories,''
  Nucl.\ Phys.\  B {\bf 579}, 101 (2000)
  [arXiv:hep-th/0001121].
}

\lref\Grisaru{
  M.~T.~Grisaru, M.~Rocek and R.~von Unge,
  ``Effective K\"ahler Potentials,''
  Phys.\ Lett.\  B {\bf 383}, 415 (1996)
  [arXiv:hep-th/9605149].
}

\lref\ChengMW{
  H.~C.~Cheng, W.~C.~Huang, I.~Low and A.~Menon,
  ``Goldstini as the decaying dark matter,''
  arXiv:1012.5300 [hep-ph].
}

\lref\DreinerTW{
  H.~K.~Dreiner, H.~E.~Haber and S.~P.~Martin,
  ``Two-component spinor techniques and Feynman rules for quantum field theory
  and supersymmetry,''
  Phys.\ Rept.\  {\bf 494}, 1 (2010)
  [arXiv:0812.1594 [hep-ph]].
}

\lref\SamuelUH{
  S.~Samuel and J.~Wess,
  ``A Superfield Formulation Of The Nonlinear Realization Of Supersymmetry And
  Its Coupling To Supergravity,''
  Nucl.\ Phys.\  B {\bf 221}, 153 (1983).
}

\lref\BrignolePE{
  A.~Brignole, F.~Feruglio and F.~Zwirner,
  ``On the effective interactions of a light gravitino with matter  fermions,''
  JHEP {\bf 9711}, 001 (1997)
  [arXiv:hep-th/9709111].
}

\lref\LutyNP{
  M.~A.~Luty and E.~Ponton,
  ``Effective Lagrangians and light gravitino phenomenology,''
  Phys.\ Rev.\  D {\bf 57}, 4167 (1998)
  [arXiv:hep-ph/9706268].
}

\lref\KomargodskiPC{
  Z.~Komargodski and N.~Seiberg,
  ``Comments on the Fayet-Iliopoulos Term in Field Theory and Supergravity,''
  JHEP {\bf 0906}, 007 (2009)
  [arXiv:0904.1159 [hep-th]].
}

\lref\ClarkXJ{
  T.~E.~Clark and S.~T.~Love,
  ``Nonlinear realization of supersymmetry and superconformal symmetry,''
  Phys.\ Rev.\  D {\bf 70}, 105011 (2004)
  [arXiv:hep-th/0404162].
}

\lref\ClarkAA{
  T.~E.~Clark, T.~Lee, S.~T.~Love and G.~H.~Wu,
  ``On the interactions of light gravitinos,''
  Phys.\ Rev.\  D {\bf 57}, 5912 (1998)
  [arXiv:hep-ph/9712353].
}

\lref\MeadeWD{
  P.~Meade, N.~Seiberg and D.~Shih,
  ``General Gauge Mediation,''
  arXiv:0801.3278 [hep-ph].
}

\lref\ClarkBG{
  T.~E.~Clark and S.~T.~Love,
  ``The Supercurrent in supersymmetric field theories,''
  Int.\ J.\ Mod.\ Phys.\  A {\bf 11}, 2807 (1996)
  [arXiv:hep-th/9506145].
}

\lref\BrignoleFN{
  A.~Brignole, F.~Feruglio and F.~Zwirner,
  ``Aspects of spontaneously broken N = 1 global supersymmetry in the  presence
  of gauge interactions,''
  Nucl.\ Phys.\  B {\bf 501}, 332 (1997)
  [arXiv:hep-ph/9703286].
}

\lref\ClarkAW{
  T.~E.~Clark and S.~T.~Love,
  ``Goldstino couplings to matter,''
  Phys.\ Rev.\  D {\bf 54}, 5723 (1996)
  [arXiv:hep-ph/9608243].
}

\lref\KleinVU{
  M.~Klein,
  ``Couplings in pseudo-supersymmetry,''
  Phys.\ Rev.\  D {\bf 66}, 055009 (2002)
  [arXiv:hep-th/0205300].
}

\lref\BrignoleCM{
  A.~Brignole, J.~A.~Casas, J.~R.~Espinosa and I.~Navarro,
  ``Low-scale supersymmetry breaking: Effective description, electroweak
  breaking and phenomenology,''
  Nucl.\ Phys.\  B {\bf 666}, 105 (2003)
  [arXiv:hep-ph/0301121].
}

\lref\DineII{
  M.~Dine, R.~Kitano, A.~Morisse and Y.~Shirman,
  ``Moduli decays and gravitinos,''
  Phys.\ Rev.\  D {\bf 73}, 123518 (2006)
  [arXiv:hep-ph/0604140].
}

\lref\AntoniadisUK{
  I.~Antoniadis and M.~Tuckmantel,
  ``Non-linear supersymmetry and intersecting D-branes,''
  Nucl.\ Phys.\  B {\bf 697}, 3 (2004)
  [arXiv:hep-th/0406010].
}

\lref\AntoniadisSE{
  I.~Antoniadis, M.~Tuckmantel and F.~Zwirner,
  ``Phenomenology of a leptonic goldstino and invisible Higgs boson decays,''
  Nucl.\ Phys.\  B {\bf 707}, 215 (2005)
  [arXiv:hep-ph/0410165].
}

\lref\GraesserBU{
  M.~L.~Graesser, R.~Kitano and M.~Kurachi,
  ``Higgsinoless Supersymmetry and Hidden Gravity,''
  JHEP {\bf 0910}, 077 (2009)
  [arXiv:0907.2988 [hep-ph]].
}
\lref\AlvarezGaumeRT{
  L.~Alvarez-Gaume, C.~Gomez and R.~Jimenez,
  ``Minimal Inflation,''
  Phys.\ Lett.\  B {\bf 690}, 68 (2010)
  [arXiv:1001.0010 [hep-th]].
}

\lref\AntoniadisNJ{
  I.~Antoniadis and M.~Buican,
  ``Goldstinos, Supercurrents and Metastable SUSY Breaking in N=2
  Supersymmetric Gauge Theories,''
  arXiv:1005.3012 [hep-th].
}

\lref\AntoniadisHS{
  I.~Antoniadis, E.~Dudas, D.~M.~Ghilencea and P.~Tziveloglou,
  ``Non-linear MSSM,''
  Nucl.\ Phys.\  B {\bf 841}, 157 (2010)
  [arXiv:1006.1662 [hep-ph]].
}

\lref\BuicanWS{
  M.~Buican, P.~Meade, N.~Seiberg and D.~Shih,
  ``Exploring General Gauge Mediation,''
  JHEP {\bf 0903}, 016 (2009)
  [arXiv:0812.3668 [hep-ph]].
}

\lref\BuicanVV{
  M.~Buican and Z.~Komargodski,
  ``Soft Terms from Broken Symmetries,''
  JHEP {\bf 1002}, 005 (2010)
  [arXiv:0909.4824 [hep-ph]].
}

\lref\ArvanitakiHQ{
  A.~Arvanitaki, S.~Dimopoulos, S.~Dubovsky  et al.,
  ``Astrophysical Probes of Unification,'' Phys.\ Rev.\  {\bf D79}, 105022 (2009).[arXiv:0812.2075 [hep-ph]].
   }

\lref\CheungMC{
  C.~Cheung, Y.~Nomura and J.~Thaler,
 ``Goldstini,''
  JHEP {\bf 1003}, 073 (2010)
  [arXiv:1002.1967 [hep-ph]];
  }

\lref\KomargodskiPC{
  Z.~Komargodski and N.~Seiberg,
  ``Comments on the Fayet-Iliopoulos Term in Field Theory and Supergravity,''
  JHEP {\bf 0906}, 007 (2009)
  [arXiv:0904.1159 [hep-th]].
}

\lref\GraesserBU{
  M.~L.~Graesser, R.~Kitano and M.~Kurachi,
  ``Higgsinoless Supersymmetry and Hidden Gravity,''
  JHEP {\bf 0910}, 077 (2009)
  [arXiv:0907.2988 [hep-ph]].
}

\lref\AntoniadisNJ{
  I.~Antoniadis and M.~Buican,
  ``Goldstinos, Supercurrents and Metastable SUSY Breaking in N=2
  Supersymmetric Gauge Theories,''
  arXiv:1005.3012 [hep-th].
}

\lref\AntoniadisHS{
  I.~Antoniadis, E.~Dudas, D.~M.~Ghilencea and P.~Tziveloglou,
  ``Non-linear MSSM,''
  Nucl.\ Phys.\  B {\bf 841}, 157 (2010)
  [arXiv:1006.1662 [hep-ph]].
}

\lref\DineSW{
  M.~Dine, G.~Festuccia and Z.~Komargodski,
  ``A Bound on the Superpotential,''
  JHEP {\bf 1003}, 011 (2010)
  [arXiv:0910.2527 [hep-th]].
}

\lref\KachruEM{
  S.~Kachru, J.~McGreevy and P.~Svrcek,
  ``Bounds on masses of bulk fields in string compactifications,''
  JHEP {\bf 0604}, 023 (2006)
  [arXiv:hep-th/0601111].
}

\lref\BuicanWS{
  M.~Buican, P.~Meade, N.~Seiberg and D.~Shih,
  ``Exploring General Gauge Mediation,''
  JHEP {\bf 0903}, 016 (2009)
  [arXiv:0812.3668 [hep-ph]].
}

\lref\DumitrescuHA{
  T.~T.~Dumitrescu, Z.~Komargodski, N.~Seiberg and D.~Shih,
  ``General Messenger Gauge Mediation,''
  JHEP {\bf 1005}, 096 (2010)
  [arXiv:1003.2661 [hep-ph]].
}

\lref\CasalbuoniXH{
  R.~Casalbuoni, S.~De Curtis, D.~Dominici, F.~Feruglio and R.~Gatto,
  ``Nonlinear Realization of Supersymmetry Algebra from
  Supersymmetric Constraint,''
  Phys.\ Lett.\  B {\bf 220}, 569 (1989).
}

\lref\IzawaHI{
  K.~-I.~Izawa, Y.~Nakai, T.~Shimomura,
  ``Higgs Portal to Visible Supersymmetry Breaking,''
[arXiv:1101.4633 [hep-ph]].
}

\lref\BenakliZZA{
  K.~Benakli, C.~Moura,
  ``Brane-Worlds Pseudo-Goldstinos,''
Nucl.\ Phys.\  {\bf B791}, 125-163 (2008). [arXiv:0706.3127
[hep-th]].
}

\lref\BuchbinderIW{
  I.~L.~Buchbinder, S.~Kuzenko, Z.~.Yarevskaya,
  ``Supersymmetric effective potential: Superfield approach,''
Nucl.\ Phys.\  {\bf B411}, 665-692 (1994). }


\Title{ } {\vbox{\centerline{Pseudo-Goldstini in
Field Theory} }}
\smallskip

\centerline{\it Riccardo Argurio,${\,^{(a)}}$ Zohar Komargodski,${\,^{(b)}}$ and Alberto Mariotti${\,^{(c)}}$}
\bigskip
\centerline{${^{(a)}}$ Physique Th\'eorique et Math\'ematique and International Solvay Institutes   }
\centerline{Universit\'e Libre de Bruxelles, C.P. 231, 1050 Bruxelles, Belgium}
\bigskip
\centerline{${^{(b)}}$ School of Natural Sciences, Institute for Advanced Study}
\centerline{Einstein Drive, Princeton, NJ 08540, USA}
\bigskip
\centerline{${^{(c)}}$ Theoretische Natuurkunde and International Solvay Institutes}
\centerline{Vrije Universiteit Brussels, Pleinlaan 2, B-1050 Brussels, Belgium}

\smallskip

\vglue .3cm

\noindent

We consider two SUSY-breaking hidden sectors
which decouple when their respective
couplings to the visible particles are switched off.
In such a scenario one expects to find two light fermions: the
Goldstino and the pseudo-Goldstino. While the former remains
massless in the rigid limit, the latter becomes massive due to
radiative effects which we analyze from several different points
of view. This analysis is greatly facilitated by a version of the
Goldberger-Treiman relation, which allows us to write a universal
non-perturbative formula for the mass. We carry out the analysis in
detail in the context of gauge mediation, where we find that the
pseudo-Goldstino mass is at least around the GeV scale and can be easily at the electroweak range, even
in low scale models. This leads to interesting and unconventional
possibilities in collider physics and it also has potential
applications in cosmology.

\Date{2/2011}

\newsec{Introduction} In this note we consider
models with multiple supersymmetry-breaking sectors. We assume
these SUSY-breaking sectors communicate only through their
respective couplings to the Supersymmetric Standard Model (SSM).
In other words, the SUSY-breaking sectors decouple when their
respective couplings to the SSM are set to zero. Such models could
naturally appear in string theory, where there may be several
independent sources of supersymmetry breaking. They may also arise
naturally in the study of quiver gauge theories. Our main
objective is to study the various field-theoretic effects that are
relevant in such a setup.

One may wonder whether having such SUSY-breaking sectors which
interact only indirectly through the SSM is natural. Indeed, in
field theory this can be perfectly natural since renormalizable
contact terms may be forbidden by gauge invariance or global
symmetries. (By contrast, decoupling in supergravity is a much
more delicate question that we will not say anything new about.)

At zeroth order in the interactions with the SSM, there are
obviously many massless Goldstini particles. Turning on the small
couplings to the SSM, one linear combination, the true Goldstino,
remains massless, while the other linear combinations get masses
from tree-level and radiative effects.

We will use several methods to analyze these corrections. We first
study the problem using the universal chiral Lagrangian for
spontaneously broken supersymmetry. The chiral Lagrangian approach
shows that the contribution from deep low momenta is quadratically
sensitive to the cutoff $\Lambda_{UV}$ of the chiral Lagrangian
\eqn\mainintro{m_{G'}\sim {1\over 16\pi^2}{m_{gaugino}^3\over
f^2}\Lambda_{UV}^2~.} Hence, the contribution is not dominated by
parameterically small momenta and one has to invoke the detailed
microscopic physics to determine the mass.
The chiral Lagrangian also shows that~\mainintro\
dominates over tree-level contributions that arise due to
electroweak symmetry breaking.

As an example of a microscopically well-defined setup we analyze in
detail two hidden sectors which only communicate with the SSM via gauge interactions. In this case we find (to all orders in the hidden sector but to leading order in the gauge coupling)
\eqn\mainintroi{m_{G'}={g^4\over 2}\left({1\over (f^A)^2}+{1\over (f^B)^2}\right)\int {d^4p\over (2\pi)^4}B_{1/2}^A(p^2)\left(C_0^B-4C_{1/2}^B+3C_{1}^B\right)(p^2)+A\leftrightarrow B~,}
where $B_{1/2}^{A,B},C_0^{A,B},C_{1/2}^{A,B},C_1^{A,B}$ are defined
through two-point correlation functions of the linear multiplets
associated to the hidden sectors $A$,$B$. These functions coincide
with the functions defined in General Gauge Mediation (GGM)~\MeadeWD. In order to show that~\mainintroi\ is indeed correct, and that the pseudo-Goldstino mass only depends on the functions that appear in GGM, we derive a generalized version of the Goldberger-Treiman relation.

If the two sectors have a common messenger scale and comparable
SUSY-breaking scales, one can roughly estimate~\mainintroi\ as
$\sim$1 GeV. On the other hand, we may consider, for instance,
different SUSY-breaking scales for the two sectors,
then~\mainintroi\ can be easily as high as $\sim$100~GeV.

We would like to elaborate more on the regime of validity of our analysis. From~\mainintroi\ it follows that our
field theory effects surely dominate over gravity as long as $m_{3/2}\sim
F/M_{PL}$ is smaller than a GeV or so. This means $\sqrt f\leq
10^9$, which covers in entirety the parameter space of
models based on gauge mediation and variations thereof. On the other
hand, since the field theoretic effects can be easily as large as 100
GeV, it may in fact be important to take them into account even in the regime of gravity mediation.

Having such heavy Goldstino-like particles in controllable low
scale models potentially leads to unconventional signatures in
collider physics and cosmology. Decays of SSM particles sometimes
proceed predominantly into the pseudo-Goldstino and may or may not
be accompanied by displaced vertices. In addition, the
pseudo-Goldstino has three-body decays with observationally interesting time scales.

A recent inspiring paper~\CheungMC\ (for earlier literature on the
subject see~\BenakliZZA) considers situations where the
gravitational effects are significant. Consistency of SUGRA
Lagrangians demands the existence of universal non-renormalizable
contact terms mixing the various sectors. Assuming that this is
the only source for mixing between the sectors, the authors
of~\CheungMC\ computed the supergravity contribution to the mass
of the pseudo-Goldstino. They found that the induced mass is $2
m_{3/2}$. Possible corrections to this result have been studied
in~\CraigYF\ and various interesting applications and variations
of this scenario are discussed in~\refs{\CheungQF,\ChengMW,
\IzawaHI}. In this note we consider theories in the rigid limit,
where these supergravity corrections are negligible.

The outline of our note is as follows. In section~2 we define
more precisely the setup, briefly review some necessary
background material about chiral Lagrangians, and discuss in detail contributions to the mass from the deep IR. In
section~3 we focus on the scenario where the hidden sectors only communicate with the visible sector via gauge interactions, and present the derivation of~\mainintroi.
In section 4 we comment on possible applications to collider physics and cosmology. Two appendices contain technical details that complement the main discussion.

\newsec{Contributions to the Pseudo-Goldstino Mass from Low Energies}

\subsec{Setup and Review}

For simplicity, and without much loss of generality,
we will henceforth restrict ourselves to two separate hidden sectors, each
communicating with the visible sector in some unspecified way. These interactions with the visible sector could be like in gauge mediation, they can include gauge messengers, Yukawa interactions with SSM fields, some mechanism responsible for $\mu-B_\mu$, and so on.

The two hidden sectors {\it decouple} if the
interactions with the SSM are switched off. Since we neglect
supergravity considerations, this form of decoupling can easily
be rendered natural due to selection rules imposed by gauge and global symmetries. (On the other hand, decoupling in the full
supergravity theory may be more problematic due to moduli; for some general arguments see, for instance,~\refs{\KachruEM,\KomargodskiRB}.)
Therefore, our zeroth order problem consists of two utterly decoupled rigid SUSY-breaking theories. Each one of them leads to a massless Goldstino. Let us now review a way of describing the Goldstino theory.

The couplings at low energy of the Goldstino to itself and to
other possibly light fields are governed by demanding invariance
under non-linearly realized supersymmetry. In general, to write
such Lagrangians we only need to insure on-shell invariance under
non-linear supersymmetry transformations. However, it is much more
convenient to have an off-shell formalism. For this reason we will
now briefly review the approach
of~\refs{\CasalbuoniXH,\nonlinear}. (There is an extensive list of
alternative approaches, e.g.~\SamuelUH\ presents several points of
view and in~\refs{\ClarkAW\BrignoleFN
\LutyNP\BrignolePE\KleinVU-\BrignoleCM} one can find later
discussions. In addition, there are different applications of
these ideas, for instance,~\refs{\AntoniadisUK\AntoniadisSE
\DineII\GraesserBU\DineSW\AlvarezGaumeRT\AntoniadisNJ-\AntoniadisHS}
is a sample of some relatively recent activity.)

Consider a general chiral superfield $\bar D_{\dot\alpha} X=0$.
The superfield $X$ includes a complex scalar, Weyl fermion, and a
complex auxiliary field. We can impose an equation
\eqn\nilp{X^2=0~.} The (nontrivial) solution is denoted $X_{NL}$
and it is given by \eqn\nonlinearsup{X_{NL}={G^2\over
2F}+\sqrt2G\theta+\theta^2F~.} Here the fields are functions of
$y=x+i\theta\sigma\bar \theta$. \foot{We are using the conventions
of~\WessCP\ everywhere in this note.} Since $X_{NL}$ is a chiral
superfield, it can be used to write arbitrary off-shell
supersymmetric actions via the usual ${\cal N}=1$ superspace. The
simplest possible theory is \eqn\AVtheory{{\cal L}=\int d^4\theta
X^\dagger_{NL}X_{NL}+\left(\int d^2\theta fX_{NL}+c.c\right)~.}
Due to~\nilp, this is also the unique theory without (superspace)
derivatives. (This has been explicitly shown to be equivalent to
the Akulov-Volkov theory. See for
example~\refs{\LiuSK\ZheltukhinXR-\KuzenkoEF}.) This can be easily
extended to include couplings to other light fields, whether they
come in complete super-multiplets or not.

Let us assume, for simplicity, that by the time we reach the soft
scale $E_{soft}$ (say between 100 GeV and a few TeV), except for
the Goldstino, no particles from the hidden sector remain. We can
read off the leading low energy couplings between SSM superfields
and the Goldstino superfield from various interaction terms in
superspace. For example, the gaugino soft-term leads to the
following couplings between the observable vector multiplet and
the Goldstino \eqn\allsoft{{\cal L}\supset {m_{\lambda}\over
2f}\int d^2\theta X_{NL}W_\alpha^2+c.c.~.} This Lagrangian is
valid at the energy scale $E_{soft}$, and is cut off at one of the
scales of the hidden sector. In models of weakly coupled
supersymmetry breaking the cut off would generically be at the
mass scale of the sGoldstino. There is a parameterically sizeable
window of energies where~\allsoft\ is valid, and higher derivative
corrections can be dropped.

Note that the one-Goldstino couplings in~\allsoft\ are not
derivative couplings. However,~\allsoft\ is equivalent to the more
familiar description of the Goldstino coupling derivatively to the
supercurrent $\sim \del^\mu G^\alpha S_{\mu\alpha}$. This equivalence can be established by performing a change of variables (which is tantamount to using the equations of motion).

\subsec{The Chiral Lagrangian}

At zeroth order in the interactions with the SSM, we have a theory of two decoupled Goldstini particles originating from the two hidden sectors labeled $A$ and $B$
\eqn\twogold{
{\cal L}=\int d^4\theta
\left(
X^{A \dagger}_{NL}X^{A}_{NL}+
X^{ B \dagger}_{NL}X^{B}_{NL} \right)
+\left[\int d^2\theta
\left( f^A X^{A}_{NL}+
 f^B X^{B}_{NL}
 \right) +
c.c\right]~. } $f^{A,B}$ and $X_{NL}^{A,B}$ are the corresponding
SUSY-breaking scales and constrained superfields.\foot{We can take
the $f^{A,B}$ to be real with no loss of generality.} We now
switch on the interactions with the SSM. Both Goldstini couple to
the SSM through the couplings \allsoft\ (and other similar
couplings). However, before discussing these couplings, one could
consider adding to~\twogold\ the tree-level term
\eqn\treeterm{\int d^2\theta \ m X^{A}_{NL} X^{B}_{NL}+c.c.~.}
Such a term can be generated by integrating-out the high momentum
modes (which one drops from the effective action), and in general
no symmetries can be used to set~\treeterm\ to zero. Note
that~\treeterm\ generates a mass for the pseudo-Goldstino, but the
true Goldstino remains massless, as it should. We cannot compute
$m$ using the effective theory, it is an unknown input of the
(unspecified) microscopic physics. However, this does not
necessarily mean that the effective theory is unavailing; we can
still learn something about the typical momentum scales in the
problem.

Whatever the corrections induced by~\allsoft\ are, they leave the true Goldstino, $G$, massless while the pseudo-Goldstino,
$G'$, gets a mass through loops of SSM fields. Denote $f_{eff}= \sqrt{(f^A)^2+(f^B)^2}$, then the physical Goldstino and pseudo-Goldstino are given by
\eqn\combi{
f_{eff}G=f^A G^A +f^B G^B
\qquad
f_{eff}G'=-f^B G^A+f^A
G^B~.}

At leading order in the supersymmetry breaking parameters, the effective action from integrating out the SSM fields is determined by the one-loop effective K\"ahler potential
\eqn\general
{
K=X_{NL}^{A\dagger}X_{NL}^{A}+X_{NL}^{B\dagger}X_{NL}^{B}+ K_{1loop} (X_{NL}^{A},X_{NL}^{A\dagger}, X_{NL}^{B},X_{NL}^{B\dagger})~.}
Once $K_{1loop}$ has been computed, one substitutes the $F$-term VEVs for $X_{NL}^{A,B}$ in order to extract the non-supersymmetric fermionic masses.
Observe that in $K_{1loop}$ each $X_{NL}$ appears at most linearly
because of~\nilp.

The pseudo-Goldstino mass can be extracted from cubic operators of the form $X_{NL}^{A}X_{NL}^{B} X_{NL}^{A\dagger}$ and other operators alike.
More precisely, the pseudo-Goldstino mass is given by
\eqn\generalmass
{m_{G'}=
{f_{eff}^2\over f^A f^B }
\left(
f^B K_{AB\bar B}+ f^A
K_{AB\bar A}
\right).
}

Consider two hidden sectors contributing in some way to the soft gaugino mass
(assuming a $U(1)$ vector superfield for notational simplicity)
\eqn\gauginotwosec{\CL= {m_{\lambda} \over 2}\int d^2\theta \left(
{\alpha^A \over f^A }X_{NL}^{A}+ {\alpha^B \over
f^B}X_{NL}^{B}\right)W_\alpha^2~.} Note that $\alpha^A+\alpha^B=1$ by definition of $m_\lambda$.

The coupling~\gauginotwosec\ is non-renormalizable, hence we need a
generalization of the one-loop K\"ahler potential to non-renormalizable theories~\Brignole. In this case the effective K\"ahler potential is quadratically divergent
\eqn\kahlergaugini{ K_{1loop}=- {1\over 16 \pi^2} \Lambda_{UV}^2
\log \left( 1+ m_{\lambda} \left({\alpha^A \over f^A}
X_{NL}^{A}+{\alpha^B \over f^B} X_{NL}^{B}+c.c. \right) \right)~.}
where $\Lambda_{UV}$ is the momentum cutoff in the loop.
The resulting pseudo-Goldstino mass is then\foot{Of course, this result can also be reproduced by an explicit one-loop computation. To perform such computations correctly one must take into account the term bilinear in the Goldstino~\nonlinearsup. The contact terms stemming from it have a similar role to the seagull term in electrodynamics; they insure the real Goldstino remains massless.}
\eqn\gauginocont{m_{G'}={\alpha^A \alpha^B \over
8\pi^2}\left({f_{eff}\over f^Af^B}\right)^2m_\lambda^3\Lambda_{UV}^2~,}
and it is {\it quadratically divergent}. This quadratic divergence can be swallowed  in the renormalization of the counterterm~\treeterm.

A quadratic divergence is bad news because it clearly signifies lack of theoretical control over the exact answer. (Strictly speaking, also finite answers may be prone to corrections from the UV, but cases where the answer is divergent are more obviously UV sensitive.) In other words, the typical momentum in the loop is
parameterically larger than the soft scale and therefore the
chiral Lagrangian does not give a universal answer. The low energy chiral Lagrangian therefore merely teaches us that one needs to understand the microscopic physics in much more detail; the universal low energy vertices do not suffice.

One can nevertheless try to estimate~\gauginocont\ by thinking of $\Lambda_{UV}$ as the supersymmetric (messenger) scale $M$. For models like gauge mediation and for comparable SUSY-breaking
scales and $\alpha^A\sim\alpha^B$
\eqn\estimate{m_{G'} \sim {m_\lambda^3M^2\over
8\pi^2f^2}\sim 10\  {\rm MeV}\quad {\rm (effective~ theory)}~. }
(We have taken the gauginos to be at the TeV scale and we have used the fact that there are $\sim10$ of them in the SSM.) The estimate~\estimate\ appears to be by and large independent of the messenger scale. Again, since we do not yet have real control over the typical momentum scale of the virtual particles, the estimate~\estimate\ is only a heuristic first crack at the problem.

One can check that the deep low energy contributions from the $B_\mu$-term, tree level effects, soft non-holomorphic scalar masses, and $A$-terms are not as significant as~\estimate. Some of these facts are established in appendix~A.

\subsec{Integrating-in the sGoldstini}

In this subsection we would like to establish that the cutoff in~\gauginocont\ is not the sGoldstino mass, even though this is the natural cutoff of the chiral Lagrangian.

At the scale of the sGolsdtino particle there is no reason for
there not to be many additional resonances. Our purpose here is
not to try and write down the most general effective action at
this scale, rather, to show that generally the sGoldstini
themselves do not render the contribution~\gauginocont\ finite.

To include the sGoldstini we simply retain the bottom components
of the $X$ fields as propagating degrees of freedom. The simplest way to model this
situation is a Polonyi model with an effective K\"ahler potential that
gives a mass to the sGoldstini: \eqn\poloqua{{\cal
L}_{\rm{hidden},i} = \int d^4\theta \big( X_iX_i^\dagger - {1\over
\Lambda_i^2} X_i^2(X_i^\dagger)^2\big) +\left(\int d^2\theta f_i X_i
+c.c.\right)~,} with $X_i = x_i +\sqrt{2}\theta \psi_i
+\theta^2 F_i$ and $i=A,B$. We might as well parametrize the effective
K\"ahler potential by the pseudomoduli masses and the SUSY-breaking scales through the relations \eqn\para{ {1\over
\Lambda_i^2} = {m_{i}^2\over 4f_i^2}~.} We will assume that the
$m_{i}$ are (well) above the soft scale.

The couplings of the Goldstini superfields to the gauge sector of the SSM are the
ones in~\gauginotwosec, replacing the
nonlinear superfields by the linear ones. We can compute the
one-loop corrections to the scalar potential for the sGoldstini
and subsequently the one-loop mass generated for the pseudo-Goldstino.
At leading order in the supersymmetry breaking scale, the
one-loop effective K\"ahler includes both the
Coleman-Weinberg effective potential for $x_i$
and the fermionic masses.

The Coleman-Weinberg potential shifts the classical $\langle x_i\rangle=0$ minimum to
\eqn\minimibis { \langle x_i
\rangle=- {1\over 8 \pi^2}  {\alpha_i
m_{\lambda}^3 \over f_i m_i^2}\Lambda_{UV}^2~. }
Expanding around this new minimum we can compute all the Goldstino bilinear terms at one-loop and we get
 \eqn\Glagbis{
 {\cal L}_{ferm}={1\over16\pi^2}
 m_{\lambda}^3 \Lambda_{UV}^2
\left(\sum_{ij}{\alpha_i \alpha_j G_iG_j\over
f_i f_j }-\sum_i{\alpha_i G_i^2\over f_i^2}\right)~.
}
From this we obtain that the induced mass for the pseudo-Goldstino is exactly
as in the previous subsection~\gauginocont. Note that taking into account the shift~\minimibis\ of the vacuum expectation value is crucial for finding one massless Goldstino.

The most important conclusion to draw from this discussion is that the pseudo-Goldstino mass is still quadratically divergent even if we include the dynamics of the sGoldstini.
The cut off scale $\Lambda_{UV}$ is thus around the fundamental scale of the theory and the full microscopic theory should be determined.

In the next section we consider a fully specified microscopic setup, and show that the typical momentum is indeed around the fundamental scale. In fact, contributions from this high scale overwhelm the low energy effects we discussed in this section.

\newsec{Hidden Sectors Communicating with the SSM by Gauge Interactions}
The next step in our exploration of the various energy scales is
to consider a complete microscopic theory. The setup we opt to
focus on is depicted in Fig.1. We consider two SUSY-breaking
theories, labeled $A$ and $B$, which communicate with the SSM via
gauge interactions. More precisely, when the SSM gauge couplings
are set to zero, the sectors $A,B$ decouple from the SSM (and thus
also from each other). These decoupled theories have some global
symmetry groups in which the SSM gauge group can be embedded and
weakly gauged.

\medskip
\epsfxsize=3.5in \centerline{\epsfbox{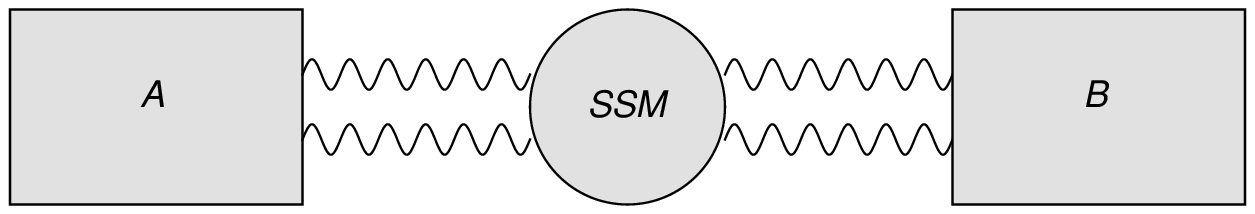}}
\centerline{Fig.1: Two SUSY-breaking theories communicating with the SSM via gauge interactions.}\medskip

In essence, this is the setup of GGM~\MeadeWD, only that the hidden sector is assumed to consist of two decoupled field theories. When the gauge couplings are turned on, the two sectors can communicate by exchanging SSM fields. Obviously, at the zeroth order in the gauge couplings, there are two exactly massless Goldstini fermions. Our goal is to find the leading nonzero contribution in an expansion in the gauge couplings.

The mass matrix for the Goldstini system, defined by $-\half
G^i{\cal M}_{ij}G^j$ with a symmetric matrix $\cal M$, is
constrained to have one zero eigenvector corresponding to the true
Goldstino. Therefore, the matrix has to be of the form
\eqn\matrixform{ {\cal M}=\left(\matrix{ -{f^B\over f^A }{\cal
M}_{AB} & {\cal M}_{AB}\cr {\cal M}_{AB} & -{f^A\over f^B }{\cal
M}_{AB} }\right)~.} Once we have calculated ${\cal M}_{AB}$, the
mass of the pseudo-Goldstino is determined  via \eqn\masspseudo{
m_{G'}=\Big( {f^B\over f^A} + {f^A\over f^B}\Big){\cal M}_{AB}~.}

Our goal is therefore to compute the first nontrivial contribution to
${\cal M}_{AB}$ in an expansion in the gauge couplings. The processes
contributing to ${\cal M}_{AB}$ consist of $G^A$ transforming into
$G^B$ via some intermediate hidden sector and SSM fields. By virtue of
the discussion in appendix A, we can neglect electroweak symmetry breaking effects. In the zeroth order in the gauge coupling $g$, we cannot have any intermediate SSM fields and thus ${\cal M}_{AB}=0$. In the next order, $g^2$, we are allowed to include one intermediate SSM fields, as schematically depicted in Fig.2. Of course this intermediate field must be the gaugino.

\medskip
\epsfxsize=2.5in \centerline{\epsfbox{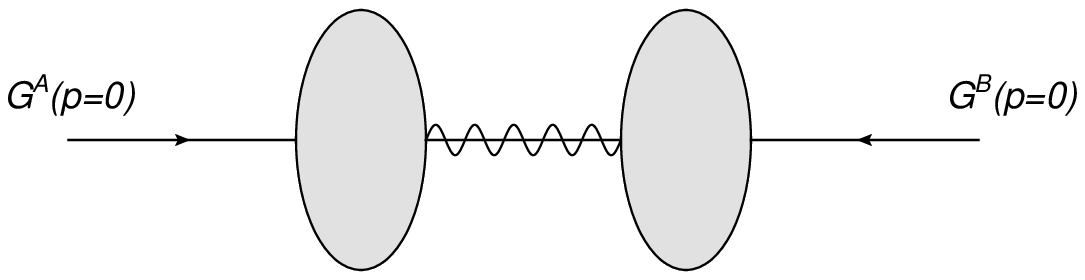}}
\centerline{Fig.2: At order $g^2$ the sectors $A$ and $B$ can communicate via one intermediate SSM gaugino.}\medskip

This process can be forbidden by messenger parity; a symmetry
which sends the vector superfield to minus itself and exchanges
all the representations accordingly. This forbids a direct zero
momentum correlation function between either of $G^{A,B}$ and a
gaugino. In fact, gauge mediation models without messenger parity
are often unappealing and we will thus assume messenger parity.

We must consider processes of order $g^4$. These allow for two intermediate SSM fields and are thus messenger parity invariant. The intermediate fields must be a gaugino and a gauge field or alternatively a gaugino and a $D$ auxiliary field. This is summarized in Fig.3.

\medskip
\epsfxsize=5.0in \centerline{\epsfbox{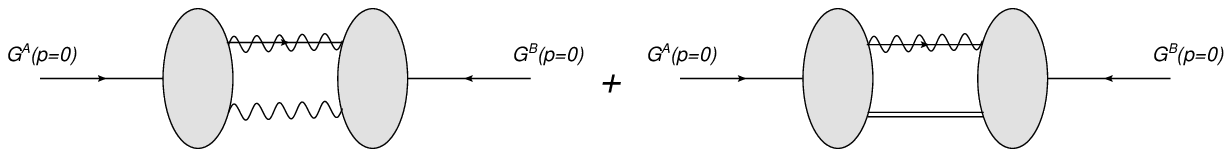}}
\centerline{Fig.3: At order $g^4$ the sectors $A$ and $B$ can communicate via two intermediate SSM fields.} \centerline{We must also add the diagrams with the gauginos flowing in the opposite direction.}\medskip

In the absence of any particular detailed knowledge of the hidden
sector we must account for the blobs formally. On the other hand, if
the theory is specified and it is weakly coupled, the blobs can be
computed in perturbation theory. For instance, in Minimal Gauge
Mediation (MGM) the blobs are, to leading order, triangles with virtual messenger fields. Therefore, the pseudo-Goldstino obtains a mass due to three-loop corrections.

In the processes of Fig.3 the external Goldstini are at zero momentum. One can therefore interpret the blobs as three-point functions of the supercurrent and two insertions of operators of the linear current multiplet. In other words, the pertinent correlation functions are of the form $\langle S_{\nu\alpha}(x) j_\mu(y) \bar j_{\alphadot}(z) \rangle$, $\langle S_{\nu\alpha}(x) J(y) \bar j_{\alphadot}(z)\rangle$, $\langle S_{\nu\alpha}(x) j_\mu(y) j_{\beta}(z)\rangle$, $\langle S_{\nu\alpha}(x) J(y) j_{\beta}(z)\rangle$. For our purposes we need the external state to be a zero-momentum Goldstino, therefore, the correlation functions above should be studied only in the limit of large $x$ (much larger than any other scale in the problem).

In this large $x$ limit the three-point functions above simplify
dramatically. The reason is that at very low energies the
supercurrent flows to the Goldstino particle
$S^{A,B}_{\mu\alpha}\sim f^{A,B}\sigma_{\mu\alpha\alphadot}\bar
G^{A,B \alphadot}$ and therefore the large $x$ limit corresponds
to inserting a zero momentum Goldstino in the correlation
function. This is the same as acting with the supercharge on the
vacuum and thus these three-point functions are related to
two-point functions of the form $\langle [\bar Q_{\dot\gamma},
j_\mu(y) \bar j_{\alphadot}(z)] \rangle$, $\langle [\bar
Q_{\dot\gamma}, J(y) \bar j_{\alphadot}(z)]\rangle$, $\langle [
\bar Q_{\dot\gamma}, j_\mu(y) j_{\beta}(z)]\rangle$, $\langle
[\bar Q_{\dot\gamma}, J(y) j_{\beta}(z)]\rangle$. These two-point
functions, in turn, appear in the calculations of soft masses in
gauge mediation. We adopt notation similar to the one in
GGM~\MeadeWD\ \eqn\hiddencor{\eqalign{&\langle
J^{A,B}(p)J(-p)^{A,B}\rangle=C_0^{A,B}(p^2)~,\cr& \langle
j_\alpha^{A,B}(p)\bar j_{\alphadot}^{A,B}(-p)\rangle =
-\sigma_{\alpha\alphadot}^\mu p_\mu C_{1/2}^{A,B}(p^2)~,\cr&
\langle j_\mu^{A,B}(p)
j_\nu^{A,B}(-p)\rangle=-\left(p^2\eta_{\mu\nu}-p_\mu
p_\nu\right)C_1^{A,B}(p^2)~,\cr& \langle j_\alpha^{A,B}(p)
j_\beta^{A,B}(-p)\rangle= \epsilon_{\alpha\beta}
B_{1/2}^{A,B}(p^2)~.   }}

The discussion above shows that the leading order contribution to
the pseudo-Goldstino mass should be captured by the functions
in~\hiddencor. A quick way to derive the precise relations between
these two-point functions and three-point functions is to start by
recalling the effective quadratic action for the vector multiplet
\eqn\effaction{\eqalign{&{1\over g^2}\CL_{eff}={1\over
2}C^A_0D^2-iC^A_{1/2}\lambda\sigma^\mu\del_\mu\bar\lambda-{1\over
4}C^A_1F_{\mu\nu}F^{\mu\nu}-{1\over
2}\left(B^A_{1/2}\lambda\lambda +c.c.\right)+A\leftrightarrow
B~.}} This breaks supersymmetry if $B_{1/2}\neq 0$ and if the $C$s
are not all equal. However, it can be supersymmetrized by adding
terms linear in the Goldstino as follows
\eqn\superGGM{\eqalign{&{1\over g^2}\CL^{one-G}_{eff}= {1\over
\sqrt2 f^A} \left(C^A_0-C^A_{1/2}\right)G^A\sigma^\mu\del_\mu\bar
\lambda D+ {i\over
\sqrt2f^A}\left(C^A_{1}-C^A_{1/2}\right)G^A\sigma_{\nu}\del_\mu\bar\lambda
F^{\mu\nu}\cr&+{iB^A_{1/2}\over\sqrt2 f^A}\left(G^A\lambda
D-{i\over 2}\lambda\sigma^\mu\bar\sigma^\nu G^A F_{\mu\nu}\right)
+A\leftrightarrow B~.}} To make the theory fully supersymmetric,
in addition to~\superGGM, we need to add terms bilinear in the
Goldstini, and terms with derivatives acting on the Goldstini. In
order to compute $\CM_{AB}$,~\superGGM\ suffices. The procedure we have invoked
here is a supersymmetric reincarnation of the Goldberger-Treiman
relation. Supersymmetric versions of the Goldberger-Treiman
relation have also been useful in the analysis of~\BaggerGM.

From here to derive the mass of the pseudo-Goldstino we only need to carry out the contractions using the vertices in~\superGGM. After the dust settles, we find that the leading order contribution to the mass of the pseudo Goldstino is
\eqn\massreadout{m_{G'}={g^4\over 2}\left({1\over (f^A)^2}+{1\over (f^B)^2}\right)\int {d^4p\over (2\pi)^4}B_{1/2}^A\left(C_0^B-4C_{1/2}^B+3C_{1}^B\right)+A\leftrightarrow B~.}
Note that the combination of the $C$ functions in the integrand is
precisely the one appearing in the formula for the soft scalar mass in
gauge mediation. The discussion in~\DumitrescuHA\ shows that
$C_0-4C_{1/2}+3C_{1}$ behaves at most like $1/p^4$ at large momentum
and it is also possible to prove that $B_{1/2}$ scales at most like
$1/p$ at large momentum.\foot{The proof of this statement goes as
follows. We start from $\langle
Q^2\left(J(x)J(0)\right)\rangle\sim \langle j_\alpha(x) j_\beta
(0)\rangle$, which was pointed out in~\BuicanWS. We can now consider
the OPE of $J(x)J(0)$. The unit operator is annihilated by
$Q^2$. Thus, the first interesting contribution in the OPE $J(x)J(0)$
is of the form $\CO/x^{4-\Delta_\CO}$, where unitarity dictates that
$\Delta_\CO>1$. Thus $\langle j_\alpha(x) j_\beta (0)\rangle$ behaves
at small $x$ like $1/x^{3-\epsilon}$ (with some positive
$\epsilon$). This implies that in momentum space $B_{1/2}(p^2)$ scales
at most like $1/p^{1+\epsilon}$ at large momentum.} Consequently, the
integral is UV convergent.\foot{One can also establish IR convergence
along the lines of~\BuicanVV.}

The computation above has been greatly simplified by the structure
of the matrix~\matrixform, which allowed us to compute $m_{G'}$
only in terms of $\CM_{AB}$. As a consistency check, we can also
compute the diagonal elements of the mass matrix $\CM_{AA}$ and
$\CM_{BB}$. In order to do this one must take into account also
the corrections to~\superGGM\ quadratic in each of the Goldstini.

We can now estimate~\massreadout\ crudely. Assume both hidden
sectors have some typical supersymmetric scale $M$ and the SUSY-breaking scales are $f^A,f^B$. To leading order in the
SUSY-breaking scales we would get \eqn\crude{m_{G'}\sim{g^4\over
(16\pi^2)^3}\left({f^A\over f^B}+{f^B\over
f^A}\right)\left({f^A\over M}+{f^B\over M}\right)~.} If the two
SUSY-breaking scales are comparable this leads to the estimate
\eqn\crudest{m_{G'} \sim {g^4\over (16\pi^2)^3}{f\over M} \sim
{g^2\over (16\pi^2)^2} m_{soft} \sim 1\ {\rm GeV}~.} (We have
included a factor of $\CO(10)$ due to the sum over the gauge
sector of the SSM.) Note that this is larger than the estimate we
obtained in the effective theory \estimate.

However we
can also entertain other possibilities. For instance, consider a
situation where the fundamental supersymmetric scales in the two
sectors are comparable but the SUSY-breaking scales are different.
To be concrete we assume that $f^A\gg f^B$ (the soft parameters
thus mostly originate in sector $A$). In this case, the
formula~\crude\ predicts an enhancement of $m_{G'}$ by $f^A/f^B$.
This ratio, however, cannot be arbitrarily large because at some
point the backreaction of the SSM on the hidden sector $B$ becomes
too large and our formalism breaks down.\foot{This is similar to
the breakdown of GGM if at zero gauge coupling there is no stable
vacuum in the hidden sector.} By computing the sGoldstino VEV in
sector $B$, we can estimate  that the backreaction is surely tame
for $f^A/f^B\ll 10^3$. (For this estimate we have assumed the mass
of the sGoldstini is around $f^{A,B}/M$.) Thus, we can easily
imagine the pseudo-Goldstino picking a mass at the electroweak
range. Note that such a (perhaps surprisingly) large mass for the
pseudo-Goldstino is achieved effortlessly and ubiquitously in low
scale models, where corrections from supergravity are completely
negligible.

One can also evaluate~\massreadout\ explicitly in a
variety of simple realizations of gauge mediation. In appendix B we consider one such example, where the two hidden sectors are copies of MGM.

\newsec{Phenomenology of Goldstini}

In the scenario presented in this note, the pseudo-Goldstino is
generically the Next-to-Lightest Supersymmetric Particle (NLSP),
with the LSP being of course the very light gravitino. The
pseudo-Goldstino is not stable and its decay can be analyzed via
the chiral Lagrangian. For instance, the terms responsible for the
gaugino mass~\gauginotwosec\ give rise to vertices of the form
$\sim G\sigma_\mu\bar\sigma_\nu \lambda F^{\mu\nu}$ which induce
three-body decays of the pseudo-Goldstino into two photons and the
true Goldstino. There are also some very important vertices with
two Goldstini. In fact, the naive estimate based on dimensional
analysis fails due to an exact cancelation between the different
vertices. An analogous story takes place in the couplings to the
SM fermions. One is left with the following estimate of the decay
width into two standard model fermions and the true
Goldstino~\ChengMW\ \eqn\widthtotal{\Gamma_{G'\rightarrow Gf\bar
f}\sim {m_{G'}^9\over 10^5f_{eff}^4}\left({(m^A_{\tilde
f})^2\tan\theta-(m^B_{\tilde f})^2\cot\theta\over m_{\tilde
f}^2}\right)^2~.} We denote $\tan\theta=f^B/f^A$ and
$(m^{A,B}_{\tilde f})^2$ are the contributions to the mass of the
slepton from the two hidden sectors, such that $(m^A_{\tilde
f})^2+(m^B_{\tilde f})^2=m_{\tilde f}^2$. There is a similar width
to decay into two photons and the true Goldstino.

Consider theories with two general SUSY-breaking scales $f^A\geq f^B$. Assuming again, for simplicity, that the messenger scales in the two sectors are comparable and taking  $m^{A,B}_{\tilde f}\sim f^{A,B}/M$ we find
\eqn\estimatetau{\tau\sim 10^{21}  \ {\rm sec}\left({f_{eff}\over 10^{10}\ {\rm GeV^2}}\right)^{4}\left({f^B\over f^A}\right)^7~.}
To derive the estimate above we have taken the mass of the pseudo Goldstino to be $m_{G'}
\sim f^A/f^B$~GeV. This gives rise to many different possibilities. For instance, when the pseudo-Goldstino is around the weak or TeV scale (i.e. $f^A/f^B\sim 10^{2-3}$) models of low scale mediation $\sqrt f\sim 10^{4-5}$ GeV give a lifetime of the order of a few seconds.
Still keeping the pseudo-Goldstino at the weak-TeV scale, we can also choose $\sqrt {f_{eff}}\sim 10^8$ GeV which leads to lifetimes of the order of $10^{23-24}$ secs. Both of these time scales have potentially interesting observable consequences~\ArvanitakiHQ.
One can of course imagine many other scenarios stemming from~\estimatetau, including scenarios with lighter pseudo-Goldstino.

One can also easily imagine many unconventional collider
manifestations of the setup here.\foot{We thank A.~Katz for
discussions on this topic.} One obvious consequence of having two
different hidden sectors is that the relation between the decay
time of the Lightest Observable-sector Supersymmetric Particle
(LOSP) and the scale of SUSY breaking is no longer universally
determined when the decay is at least for a significant fraction
to the pseudo-Goldstino. This can have several different
consequences.

For instance, consider two hidden sectors with comparable
messenger scales but with a possible hierarchy in the SUSY
breaking scales. From the couplings~\gauginotwosec\ we see that
the gaugino is equally likely to decay to either of the Goldstini
(since the dependence on $f$ cancels and only the supersymmetric
scale remains). Therefore, if the LOSP is bino- or wino-like,
and it is heavier than the pseudo-Goldstino, many of the processes
of the SSM will terminate in a heavy, long lived, pseudo-Goldstino
(the decay can be prompt or there can be displaced vertices). This
also comes accompanied by an isolated photon from the last step of
the decay. Having such an invisible heavy particle as missing
energy is clearly different from conventional scenarios of gauge
mediation where the missing energy is carried away by practically
massless objects. It is also distinguishable from gravity
mediation, where the LOSP is stable on collider time scales and
therefore, if it is a gaugino, no isolated photons are expected.

We will not attempt to classify all the scenarios and signatures
here. The very brief remarks above are just to demonstrate that
unusual collider and cosmological signatures are definitely
possible. Clearly, it will be interesting to investigate the
various possibilities further. It is also important to study more
general hidden sector paradigms, beyond gauge mediation.

\bigskip
\bigskip
 \noindent {\bf Acknowledgments:}
\nobreak We would like to thank M.~Bertolini, M.~Buican,
T.~Dumitrescu, G.~Ferretti, A.~Katz, K.~Mawatari, N.~Seiberg and
D.~Shih for helpful conversations. R.A. is a Research Associate of
the Fonds de la Recherche Scientifique--F.N.R.S. (Belgium). The
research of R.A. is supported in part by IISN-Belgium (conventions
4.4511.06, 4.4505.86 and 4.4514.08) and by the ``Communaut\'e
Fran\c{c}aise de Belgique" through the ARC program. A.M. is a
Postdoctoral Researcher of FWO-Vlaanderen. A.M. is also supported
in part by FWO-Vlaanderen through project G.0114.10N. R.A. and
A.M. are supported in part by the Belgian Federal Science Policy
Office through the Interuniversity Attraction Pole IAP VI/11. Z.K.
gratefully acknowledges support by DOE grant DE-FG02-90ER40542.
Any opinions, findings, and conclusions or recommendations
expressed in this material are those of the authors and do not
necessarily reflect the views of the National Science Foundation.

\appendix{A}{Other Contributions from Low Energies}
\subsec{The $B_\mu$-term}

The $B_\mu$-term leads to the following coupling between the
observable  Higgs multiplets and the Goldstino \eqn\allsoft{{\cal
L}\supset {B_\mu\over f}\int d^2\theta X_{NL}H_uH_d+c.c.~.} In the
case at hand, we have two different SUSY-breaking sectors, each
contributing independently to the $B_\mu$-term \eqn\twocont{\int
d^2\theta B_\mu\left( {\lambda^A \over f^A}
X_{NL}^{A}H_uH_d+{\lambda^B \over f^B} X_{NL}^{B}H_uH_d
\right)+\int d^2\theta \mu H_uH_d+c.c.~.} Of course
$\lambda^A+\lambda^B=1$ since $B_\mu$ is the actual physical soft
term. We also included the $\mu$-term in~\twocont.

The effective K\"ahler potential from integrating out the SSM
Higgs fields~\twocont\ (which for our purpose appear only
quadratically) is~\refs{\BuchbinderIW,\Grisaru} (we keep only the
significant terms) \eqn\koneloop { K_{1loop}=-{1\over 32 \pi^2}
Tr\left[{\cal M}{\cal M}^\dagger \log {{\cal M}{\cal M}^\dagger
\over \Lambda^2}\right]~, } where \eqn\matrixoneloop{{\cal
M}=B_\mu \left(\matrix{0&{\lambda^A \over f^A} X_{NL}^{A}
+{\lambda^B \over f^B} X_{NL}^{B}+{\mu\over B_\mu} \cr {\lambda^A
\over f^A} X_{NL}^{A} +{\lambda^B \over f^B} X_{NL}^{B}+{\mu\over
B_\mu}& 0 } \right)~.} The resulting contribution to the mass of
the pseudo-Goldstino is \eqn\massfake{m_{G'}={\lambda^A \lambda^B
\over 16\pi^2}\left({f_{eff}\over f^Af^B}\right)^2{B_\mu^3\over
\mu}~.} The result of this calculation in the chiral Lagrangian is
finite.

If the $B_\mu$-term is around the soft scale, and SUSY breaking
occurs at low scales ($f\sim 10^8$ GeV$^2$),~\massfake\ is of the
order of $10$ eV but it can increase to as much as an MeV if
$B_\mu$ is larger by an order of magnitude. This is numerically
smaller than the effect from low momentum vector multiplet
loops~\estimate. The typical momentum of the virtual particles
is low and the result of the low energy calculation is convergent.
This however, does not necessarily mean the result~\massfake\ is
reliable; there could still be contributions from the UV.

\subsec{Tree-Level Contributions}

Effects of electroweak symmetry breaking induce tree-level
contributions to the pseudo-Goldstino mass. For instance, due to
electroweak symmetry breaking, the hypercharge $D$-term is generally
nonzero. Let us denote its value simply by $D$. Then, the
gaugino soft mass terms~\gauginotwosec\
lead to the following
mass matrix in the space of the three fermions $G^A,G^B,\lambda$
\eqn\massmatrix{{\cal M}=m_\lambda\left(\matrix{{\alpha^A D^2\over
2(f^A)^2} & 0 & {\alpha^A D\over \sqrt2 f^A}\cr 0 & {\alpha^B
D^2\over 2(f^B)^2} & {\alpha^B D\over  \sqrt2f^B} \cr
  {\alpha^A D\over \sqrt2 f^A}  & {\alpha^B D\over \sqrt2 f^B} & 1}\right)~.}
If the hypercharge $D$-term is set to zero, the only tree-level contribution that remains is the gaugino mass term and the pseudo-Goldstino is massless at tree-level. In this case only radiative effects exist.
The matrix~\massmatrix\ has, schematically (assuming the two SUSY-breaking
scales are comparable and neglecting corrections in $D/f$), the eigenvalues \eqn\eigenlist{\left(0,
{m_\lambda D^2\over
f^2},m_\lambda\right)~,}
corresponding to the real Goldstino, the pseudo-Goldstino, and the
gaugino, respectively. Therefore, the tree-level contribution to
the mass of the pseudo-Goldstino scales like
\eqn\answe{m_{G'}^{(tree)}\sim m_\lambda {D^2\over f^2}~,} which
is parameterically smaller than the low-energy quantum effect~\estimate. For
instance,~\answe\ is maximized for low scale models
($\sqrt f\sim 10$ TeV), where it can be roughly estimated as
10~keV.

An additional tree-level effect could arise from the Higgs superfields picking up an $F$-term. For instance, the source for the $B_\mu$ term~\twocont\ also induces, upon electroweak symmetry breaking, a tree-level mass for $G'$. With similar techniques, we can estimate it to be parameterically smaller than~\estimate.

Tree-level contributions arising from electroweak breaking are therefore negligible.

\appendix{B} {Two Copies of MGM}
Here we consider an explicit example of the class of models studied in section 3.
Suppose that each of the two hidden sectors is a copy of Minimal Gauge
Mediation (MGM).
For simplicity we discuss the case in which the SSM is represented by a $U(1)$ gauge group. The messengers $\Phi_i$, $\tilde\Phi_i$ are massive and interact with the spurions $X^i$ ($i=A,B$) as
\eqn\basiclag{\eqalign{&{\cal L}_i = \int d^2\theta \left( h_i X^i \Phi_i \tilde \Phi_i + M_i \Phi_i
\tilde \Phi_i\right) ~,}}
where the spurions $X^i$ acquire $F$-term VEVs $f^i$ and contain the
Goldstini in the $\theta$ components. We assume without loss of
generality that the bottom components $x^i$ are stabilized at the origin.

The leading contributions to  ${\cal
M}_{AB}$ are easily listed.
They consist of triangle diagrams fused by SSM fields. In each of the diagrams, only one of the triangles needs an $F$-term insertion. We summarize the various three-loop diagrams in Fig.4.

\medskip
\medskip
\epsfxsize=4.5in \centerline{\epsfbox{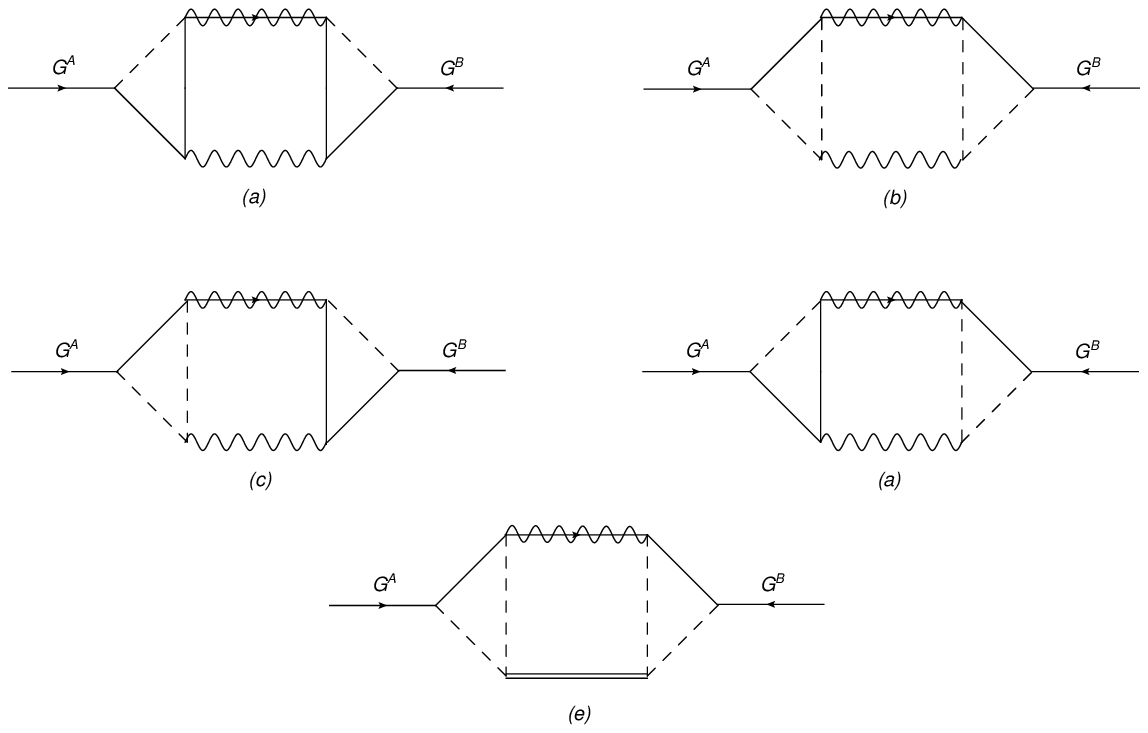}}
\centerline{Fig.4: The three-loop diagrams contributing to ${\cal M}_{AB}$.}\medskip

The computation can be done in two steps. First we compute
the momentum dependence of the triangle vertices
when the Goldstino particle is at zero momentum and the gauge-multiplet
field have some general momentum $p$.
To compute the three-loop diagrams we then have to convolve these momentum dependent
vertices with the gauge-multiplet propagators and perform the integration.
This mimics the procedure we have carried out when discussing the general
framework in section 3.

We only display the computations to first order in the supersymmetry breaking scales $f^A$ and $f^B$.
Doing the explicit computation we find that the vertices in momentum space with zero momentum Goldstini are
\eqn\vertmom{\eqalign{&{1\over g^2}\CL^{one-G}_{eff}= {iV_0^A(p^2/M_A^2) \over\sqrt2 f^A}\left(G^A\lambda D-{i\over 2}\lambda\sigma^\mu\bar\sigma^\nu G^A F_{\mu\nu}\right) \cr&+{i V_1^A(p^2/M_A^2) \over \sqrt2 f^A} G^A\sigma^\mu p_{\mu}\bar \lambda D-
{V_2^A(p^2/M_A^2)\over \sqrt2f^A}G^A\sigma_{\nu}p_{\mu} \bar\lambda F^{\mu\nu}+A\leftrightarrow B~.}}
The functions $V_0^A(p^2/M_A^2)$, $V_1^A(p^2/M_A^2)$, $V_2^A(p^2/M_A^2)$ are given by
\eqn\vertint{
\eqalign{&
V_0^A(p^2/M_A^2)=
{1 \over 16 \pi^2} {f^A \over M_A} b(p^2/M_A^2)=
{1 \over 16 \pi^2} {f_A \over M_A} \int_0^1 dy {1 \over 1+ y(1-y) x_A}~,
\cr
&
V_1^A(p^2/M_A^2)=V_2^A(p^2/M_A^2) =
{1 \over 16 \pi^2}  {f_A^2 \over M_A^4}
v(p^2/M_A^2)
=
{1 \over 16 \pi^2}  {(f^A)^2 \over M_A^4}
\int_0^1 dy { y^2(1-y) \over (1+y(1-y) x_A)^2}~,
}}
where $x_A=p^2/M_A^2$, and the $y$ integrals can be done analytically.
Comparing these with the expressions for the $B$ and $C$ functions of MGM (see~\MeadeWD)
one can verify that the momentum dependent vertices are exactly as in~\superGGM.
(Note in particular that at this order in SUSY-breaking for MGM we have $C_1=C_0$.)
The final answer for the pseudo-Goldstino mass is then
\eqn\massgoldstino{
m_{G'}={ g^4 \over  (16 \pi^2)^3}
\left(
{1 \over (f^A)^2}+ {1 \over (f^B)^2}
\right) \int_0^\infty dx\, x {f^A (f^B)^2 \over M_A} v\left(x\right) b\left({xM_B^2 \over M_A^2}\right) + A\leftrightarrow B~.
}

We can calculate this integral in different limits.
For $M_A=M_B\equiv M$ we get
\eqn\massgoldnum{
m_{G'}\simeq 4.21 \times {g^4 \over  (16 \pi^2)^3}
{((f^A)^2+(f^B)^2)(f^A+f^B) \over f^A f^B M}~,
}
in agreement with \crude.
We can also consider cases where the messenger masses are not the same. This may open up new interesting possibilities, but we leave it for the future.

 \listrefs
 \end